# Strong Room-Temperature Ferromagnetism in Co$^{2+}$-Doped TiO$_2$ made from Colloidal Nanocrystals

*J. Daniel Bryan,[†] Steve M. Heald,[‡] Scott A. Chambers,[‡] and Daniel R. Gamelin[†,*]*

Department of Chemistry, University of Washington, Seattle, WA 98195-1700, and

Fundamental Science Division, Pacific Northwest National Laboratory, P.O. Box 999, MS K8-93,

Richland, WA 99352

*To whom correspondence should be addressed: Gamelin@chem.washington.edu
[†]University of Washington, [‡]Pacific Northwest National Laboratory

*Abstract:* Colloidal cobalt-doped TiO$_2$ (anatase) nanocrystals were synthesized and studied by electronic absorption, magnetic circular dichroism, transmission electron microscopy, magnetic susceptibility, cobalt K-shell X-ray absorption spectroscopy, and extended X-ray absorption fine structure measurements. The nanocrystals were paramagnetic when isolated by surface-passivating ligands, weakly ferromagnetic ($M_s \approx 1.5 \times 10^{-3}$ $\mu_B$/Co$^{2+}$ at 300 K) when aggregated, and strongly ferromagnetic (up to $M_s = 1.9$ $\mu_B$/Co$^{2+}$ at 300 K) when spin-coated into nanocrystalline films. X-ray absorption data reveal that cobalt is in the Co$^{2+}$ oxidation state in all samples. In addition to providing strong experimental support for the existence of intrinsic ferromagnetism in cobalt-doped TiO$_2$, these results demonstrate the possibility of using colloidal TiO$_2$ diluted magnetic semiconductor nanocrystals as building blocks for assembly of ferromagnetic semiconductor nanostructures with potential spintronics applications.

# INTRODUCTION

Ferromagnetic semiconductors are envisioned as key functional components of many spin-based semiconductor devices.[1,2] If successfully produced, such spintronics devices may offer lower power consumption and greater operating speeds than their charge-based analogs, and may display new functional properties having no charge-based analog. Successful prototype devices have used nanoscale manganese-doped ZnSe,[3] GaAs,[4] and InAs[2] diluted magnetic semiconductors (DMSs) to accomplish spin polarization or spin detection, but the cryogenic Curie temperatures ($T_C$) required for ferromagnetic ordering have hindered the development of practical devices based on these DMSs. The development of new high-$T_C$ ferromagnetic semiconductors has consequently emerged as a research subject of considerable urgency.

Cobalt-doped $TiO_2$ ($Co^{2+}:TiO_2$, anatase) has attracted intense attention since the initial reports of ferromagnetism above 300 K in thin films of this material grown by pulsed laser deposition (PLD)[5] and oxygen plasma assisted molecular beam epitaxy (OPA-MBE).[6] The origin of this ferromagnetism continues to be debated in the literature, however. Some thin films have been shown to contain cobalt metal nanoparticles,[7,8] fostering the widespread belief that metallic nanoparticles are responsible for the ferromagnetism in other samples as well. Alternatively, combined magnetic force and atomic force microscopy studies of $Co^{2+}:TiO_2$ thin films grown by OPA-MBE have directly identified ferromagnetic islands of cobalt-enriched anatase $TiO_2$, results that argue for the existence of intrinsic ferromagnetism in at least some $Co^{2+}:TiO_2$ preparations.[9] Recently, the anomalous Hall effect was observed in $Co^{2+}:TiO_2$ in the rutile polymorph,[10] and this observation was used to argue for intrinsic DMS ferromagnetism in this material. The anomalous Hall effect was subsequently observed in $TiO_2$ that contained metallic cobalt nanoparticles as a separate phase,[11] however, re-igniting debate over the origins of the ferromagnetism in this material. In each of the above studies, the materials were prepared by vacuum deposition methods.

In this paper, we describe the direct chemical synthesis of colloidal $Co^{2+}$:$TiO_2$ nanocrystals (NCs) and the use of these NCs for the preparation of thin films showing very strong room-temperature ferromagnetism. A key feature of this new preparation is the use of $Co^{2+}$ ionic precursors and oxidizing reaction conditions that preclude the formation of metallic cobalt. Transition-metal impregnated or doped $TiO_2$ prepared by wet chemical synthesis has been widely studied for photocatalytic, photovoltaic, or gas sensing applications,[12] but the magnetic properties of these materials have remained largely unexplored. Direct chemical syntheses of DMS nanocrystals may offer several attractive possibilities for the study of DMSs that are complementary to those offered by vacuum deposition syntheses. Direct chemical syntheses of several colloidal II-VI,[13,14] III-V,[15] and IV-VI[16] DMS NCs have been reported recently, and in many cases these syntheses include purification steps able to eliminate surface-bound dopants or other impurities that may otherwise compromise the physical properties of the target DMSs. Such purification steps are not available for the vacuum deposited materials. In addition, direct chemical syntheses are generally amenable to scaled-up processing, allowing gram quantities of novel DMSs to be prepared relatively quickly. Finally, the solution compatibilities of the DMS NCs prepared by direct chemical synthesis offer new opportunities for nanotechnological applications, in particular as building blocks in more complex nanostructures formed by self assembly or other solution processing approaches.

EXPERIMENTAL SECTION

**A. Sample Preparation.** The synthesis of $Co^{2+}$:$TiO_2$ NCs was adapted from an inverse micelle procedure shown previously to yield colloidal anatase $TiO_2$ NCs.[17] In a typical synthesis, an inverse micelle solution (IMS) was prepared by stirring 50ml of cyclohexane, 5 g of dioctyl sulfosuccinate, sodium salt (AOT) (Aldrich 96%), 0.9 ml of $H_2O$, and 0.207 g of the $Co^{2+}$ salt $Co(NO_3)_2·6H_2O$ for several hours. Once the $Co^{2+}$ salt was fully dissolved, the solution was placed under $N_2$ for storage until needed. An anaerobic solution containing 30 ml of 1-hexanol (Aldrich 99%, vacuum distilled) and 3 ml of titanium isopropoxide (Strem, purified by recrystallization) was then slowly transferred to the Co$^{2+}$/IMS under N$_2$. The mixture was stirred for 30 minutes and then aged in air for 24 hours. An amorphous product containing Co$^{2+}$ could be isolated from the micelle solution at this stage. To induce crystallization, solutions of the amorphous product were sealed under air and heated at 150 ºC for 24 hours in a Teflon-lined stainless steel autoclave (Parr 4749). The resulting transparent tan solution contained nanocrystalline cobalt-doped TiO$_2$. Two procedures were used to isolate and purify the NCs. Procedure (1): selective precipitation and washing with ethanol. Procedure (2): addition of trioctylphosphine oxide (Aldrich, tech grade) (TOPO, 60mg/ml of solution) followed by heating at 80-95 °C for one hour, selective precipitation with acetone, and separation by centrifugation. After purification by Procedure 2, the NCs were redispersed in a small volume of toluene to yield transparent colloidal suspensions of high optical quality, showing no scattering of visible photons noticeable by eye. Thin nanocrystalline films were prepared by spin-coating colloidal TOPO-capped Co$^{2+}$:TiO$_2$ NCs onto polycrystalline sapphire substrates. Multiple coats were deposited, annealing in air at 350 ºC for ca. 1 minute between coats.

**B. Physical Methods.** X-ray powder diffraction data for nanocrystals precipitated from toluene were collected on an 800 W Philips 1830 powder diffractometer using an etched glass plate sample holder, and for thin films were collected on a 12 kW Rigaku Rotoflex rotating anode diffractometer. NaCl was used as an instrumental peak broadening standard. High resolution transmission electron microscopy (TEM) images were collected at the Pacific Northwest National Laboratory on a JEOL 2010 (200kV) microscope with a high-brightness LaB$_6$ filament electron source. Dopant concentrations were measured by inductively coupled plasma atomic emission spectroscopy (ICP-AES, Jarrel-Ash model 955) using elemental calibaration standards (High Purity Standards). Scanning electron microscopy (SEM) images were collected using a Scion XL-30 field emission microscope. Electronic absorption spectra were collected on a Cary 5E (Varian) spectrophotometer using 1 cm path-length quartz cuvettes. Magnetic circular dichroism (MCD) spectra were collected using a UV/VIS/NIR MCD instrument constructed from an AVIV 40DS spectropolarimeter and a high-field superconducting magneto-optical cryostat (Cryo-Industries SMC –1659 OVT) with a variable temperature sample compartment positioned in the Faraday configuration. Magnetic susceptibilities were measured using a Quantum Design Magnetic Property Measurement System equipped with a reciprocating sample option. Film thicknesses were determined by an Alpha Step Profilometer (Model 500, KLA-Tencor, San Jose, CA).

X-ray absorption measurements were made at the PNC-CAT beam line 20-BM at the Advanced Photon Source (Argonne National Laboratory) using a Si(111) double crystal monochromator with energy resolution of ~1.2 eV at the Co edge. Beam harmonics were rejected using a rhenium-coated mirror at 7 mrad incident angle. Co fluorescence was detected using a Ge detector. Al foil (~100 mm) was used to filter out the Ti fluorescence. Thirty scans were collected for each sample. XAS/EXAFS data fitting and analysis were performed as described previously.[18]

RESULTS AND ANALYSIS

Addition of titanium isopropoxide to the $Co^{2+}$/IMS solution resulted in a color change from rose to purple, the purple solution showing an absorption maximum at 17 640 $cm^{-1}$. A similar purple color has been reported for the oxo-bridged heterometallic complex $Co_2Ti_2(acetylacetonate)_2(isopropoxide)_{10}$ ($Abs_{max}$ at 17 700 $cm^{-1}$), formed by heating cobalt acetylacetonate with titanium isopropoxide in a nonpolar solvent.[19] Figure 1(a) shows the featureless powder X-ray diffraction observed for the purple product that could be isolated at this stage. We conclude that the mixture of these two solutions forms an amorphous oligomeric metal-oxo gel[20] into which $Co^{2+}$ ions have been randomly incorporated. We have found that careful attention must be given to incorporation of $Co^{2+}$ into this precursor, as success in this step greatly facilitates incorporation of high concentrations of $Co^{2+}$ into the $TiO_2$ nanocrystals in the following stage of the synthesis. This gel thus serves as a pre-assembled doped precursor for the preparation of doped $TiO_2$ nanocrystals.

Autoclaving the amorphous purple species causes its transformation into TiO$_2$ nanocrystals, which remain suspended in the inverted micelles and can be isolated and purified using Procedures (1) or (2) as described in the experimental section. Figure 1(b) shows powder X-ray diffraction data for a sample of TOPO-capped 3.0 ± 0.1 % Co$^{2+}$:TiO$_2$ NCs (hereafter referred to as NC1), isolated from the inverted micelles by Procedure (2). Only, the size-broadened diffraction peaks of anatase TiO$_2$ are observed. Analysis of the (101) peak widths using the Scherrer equation[21] yielded an effective NC diameter of ca. 4 nm, similar to the sizes reported originally for undoped TiO$_2$ nanocrystals (4.4 ± 0.5 nm).[17] Figure 2A shows a high-resolution TEM image of the same NCs. The NCs have dimensions consistent with those anticipated from the X-ray diffraction peak widths, but are slightly elongated along the c-axis direction. Preferential growth of pure TiO$_2$ nanocrystals along the c-direction has been reported previously,[22] and the ellipticity observed in Figure 2A indicates that it should be possible to generate similarly anisotropic nanocrystals of this DMS from solution under proper experimental conditions.

Although TiO$_2$ nanocrystals are susceptible to aggregation, the lack of noticeable visible-light scattering by the NC1 suspension suggests that little aggregation has occurred in this case. A low-resolution TEM image (Supporting Information) also did not show any obvious aggregation, although aggregation may still have occurred on some length scale. The suppression of aggregation is apparently attributable to the TOPO-capping procedure used to isolate and purify these nanocrystals (Procedure (2)).

Isolation Procedure (2) involves TOPO ligand exchange at the surfaces of the Co$^{2+}$:TiO$_2$ NCs. When the resulting TOPO-capped nanocrystals are precipitated from the IMS solution, a blue supernatant remains. The absorption spectrum of the supernatant was monitored vs. heating time to ensure that the process had run to completion. The blue color comes from solvated Co$^{2+}$ ions in tetrahedral coordination environments. We have previously observed that TOPO is capable of removing Co$^{2+}$ from the surfaces of ZnO DMS NCs by complexation,[14] also yielding tetrahedrally coordinated Co$^{2+}$ ions in solution, and the appearance of Co$^{2+}$ in the supernatant here suggests that similar surface cleaning occurs with TiO$_2$

DMS NCs as well. This conclusion is supported by quantitative analysis of the cobalt concentrations in the NCs, determined using ICP-AES. In a control experiment performed upon a single synthesis batch, the $Co^{2+}$ concentration in ca. 4 nm diameter NCs isolated using Procedure (1) was 3.96 ± 0.12 % (i.e. $Co_{0.040}Ti_{0.960}O_{2-\delta}$), whereas a sample from the same batch isolated and purified using Procedure (2) contained only 3.03 ± 0.13 % $Co^{2+}$. In spherical 4.0 nm diameter $TiO_2$ NCs, roughly 30 % of the $Ti^{4+}$ ions reside in the outermost cation shell at or near the NC surfaces (slightly more if the crystals are elongated as in this case, vide infra), and this percentage is similar to the fraction of $Co^{2+}$ lost to solution with purification by Procedure (2) (~25 %). These results suggest that the majority if not all of the surface bound $Co^{2+}$ ions are removed by Procedure (2). This synthetic procedure thus allows control over the $Co^{2+}$ speciation by selectively removing surface $Co^{2+}$. The synthesis and purification is summarized in Scheme 1. As colloids, the DMS NCs obtained from this synthesis are then suitable for solution processing methods, similar to molecular species.

**Scheme 1:**

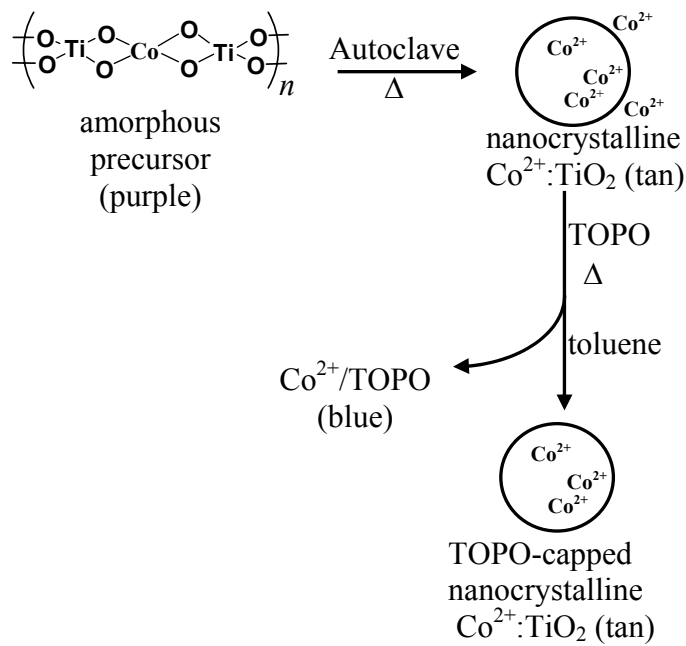

Figure 3A shows 300 K absorption spectra of colloidal $TiO_2$ NCs (undoped) and 3.0 ± 0.1 % $Co^{2+}:TiO_2$ NCs (NC1) suspended in toluene. Whereas the undoped $TiO_2$ NCs show only the band gap absorption onset of TiO$_2$ at ~25 800 cm$^{-1}$ (3.2 eV), the Co$^{2+}$:TiO$_2$ NCs show a new broad absorption tail that extends well into the visible, giving the colloids their distinct tan color (Figure 3A inset). Low-temperature absorption experiments failed to resolve any structure in this tailing absorbance feature with the exception of a poorly resolved shoulder at ca. 19 000 cm$^{-1}$ ($\varepsilon \approx 6$ M$^{-1}$ cm$^{-1}$) that can be tentatively associated with the low symmetry split $^4T_1(F) \rightarrow {}^4T_1(P)$ ligand-field transition of pseudo-octahedral Co$^{2+}$. The analogous transition in Co(H$_2$O)$_6^{2+}$ is observed at 19 400 cm$^{-1}$ ($\varepsilon = 4.8$ M$^{-1}$ cm$^{-1}$),[23] and although a similar ligand field strength is anticipated for Co$^{2+}$ in anatase TiO$_2$, it is expected to be substantially broadened in the latter by the reduction in site symmetry and by possible charge compensation inhomogeneities. The presence of this transition is confirmed by MCD spectroscopy (Figure 3B), which reveals a broad negative pseudo-A term feature in the same energy region (- 19 000 cm$^{-1}$, + 21 500 cm$^{-1}$). The MCD intensity vs. temperature and field is consistent with magnetization of a zero-field split S = 3/2 ground state. To higher energy (> ~22 500 cm$^{-1}$), positive C-term MCD intensity is observed that is associated with the broad tailing sub-bandgap absorbance in Figure 3A, and is seen from the intensity variations with field to arise from the same magnetic chromophore as the pseudo-A term feature. Its MCD and absorption intensities and its absence from the absorption spectrum of the analogous nanocrystals not containing Co$^{2+}$ (Figure 3A) indicate that this broad tailing sub-bandgap absorbance arises from a charge transfer process involving Co$^{2+}$.

Figure 4 shows (A) orbital and (B) term energy level diagrams anticipated for high-spin Co$^{2+}$ ions occupying the cation sites of anatase TiO$_2$. The anatase cation site has $D_{2d}$ site symmetry, as seen in the unit cell shown in Scheme 2 (dark spheres = Ti, light spheres = O). Relative to the octahedral high-symmetry limit, displacement of the equatorial ligands to above and below the equatorial plane splits the d-orbital degeneracies, leaving only d$_{xz}$ and d$_{yz}$ degenerate. This distortion stabilizes those d orbitals that have only x and y components, and results in the ground-state configuration shown in Figure 4A, right. This configuration yields a $^4$E term as the $D_{2d}$ (anatase) ground state (Figure 4B).

**Scheme 2**

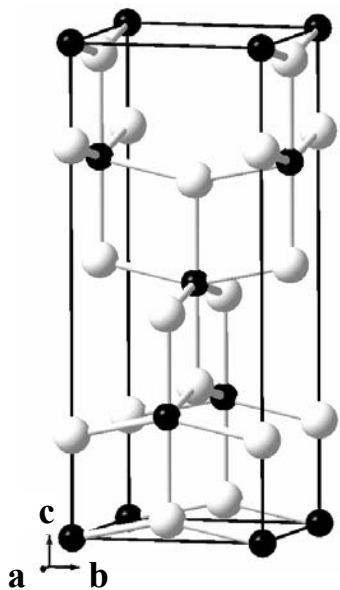

The specific energies shown in Figure 4B were estimated by calculating the effect of distortion of $Co(H_2O)_6^{2+}$ into the anatase geometry within the context of the angular overlap model.[24] The idealized ligand field parameters of $Co(H_2O)_6^{2+}$ ($Dq$ = 927 cm$^{-1}$, B = 845 cm$^{-1}$, C = 4.5B)[23] were used, approximating the oxos as pure σ donors. The resulting energies provide a useful estimate of the excited state energies anticipated for $Co^{2+}$ in anatase $TiO_2$, which has similar $Co^{2+}$-O bond lengths ($R_{avg}$ ≈ 2.07 Å in $Co(H_2O)_6^{2+}$ vs. $R_{avg}$ ≈ 2.04 Å in anatase $TiO_2$, vide infra) and in which the oxygen ligands are also three coordinate. These calculations yield the anticipated $^4E$ low-symmetry ground state, and a low-symmetry split $^4T_{1g}$ term centered at 18 724 cm$^{-1}$ above the $^4E$ ground state, in reasonable agreement with the energy of the broad absorption and MCD pseudo-A-term feature in Figure 3 (centered at ca. 20 000 cm$^{-1}$).[25] Within the above approximations, the total splitting of the $^4T_{1g}$ term is ca. 650 cm$^{-1}$. The pseudo-A-term MCD feature is thus attributed to the $^4B_1$ and $^4E$ states of $^4T_{1g}$ parentage, which gain equal and opposite MCD intensity through spin-orbit coupling.[26] In summary, the broad absorption and pseudo-A-term MCD feature centered at ca. 20 000 cm$^{-1}$ are consistent with a scenario in which the $Co^{2+}$ ions occupy the distorted 6-coordinate cation sites of anatase $TiO_2$ (Scheme 2) and possess a $^4E$ ground state. This $^4E$ ground state is then subject to zero-field splitting (not shown in Figure 4B) due to spin-orbit coupling with the nearby quartet excited states.

The quartet electronic ground state is confirmed by magnetic susceptibility measurements. Figure 5 shows the results of variable-temperature magnetic susceptibility measurements on 3.0 ± 0.1 % $Co^{2+}$:$TiO_2$ NCs (NC1). The effective magnetic moment ($\mu_{eff}$) is temperature independent between 100 and 300 K, having a value of $\mu_{eff}$ = 4.2 $\mu_B$/$Co^{2+}$ that is consistent with the value recorded for $Co^{2+}$ in bulk microcrystalline anatase $TiO_2$ ($\mu_{eff}$(300 K) = 4.1 $\mu_B$/$Co^{2+}$).[27] Although high-spin octahedral $Co^{2+}$ has a spin-only effective magnetic moment of $\mu_{eff}^{s.o.}$ = 3.87 $\mu_B$/$Co^{2+}$, the $^4T_{2g}$ ground state of octahedral $Co^{2+}$ also possesses orbital angular momentum (L' = 1). Following Kotani,[28] ligand field theory predicts that octahedral $Co^{2+}$ in the $^4T_1$ ground state with $\lambda = -\zeta_{3d}/2S = -172$ cm$^{-1}$ and A ~ 1.4 will have $\mu_{eff}$ = 5.1 $\mu_B$/$Co^{2+}$ at 300 K. 300 K effective magnetic moments of $\mu_{eff} \approx 5$ $\mu_B$/$Co^{2+}$ are typically observed in $Co^{2+}$ coordination complexes with nearly octahedral geometries.[23] The effective magnetic moment will approach the spin-only value with increasing low-symmetry splitting of the $^4T_{2g}$ ground state. In the $D_{2d}$ cation site symmetry of anatase $TiO_2$, the $O_h$ $^4T_{2g}$ term is split into $^4B$ and $^4E$ states separated by $\delta \approx 627$ cm$^{-1}$ ($|\delta/\lambda|$ = 3.6, Figure 4B), thereby largely quenching the effective orbital angular momentum of the parent $^4T_{2g}$ ground state. The 300 K effective magnetic moment of 4.2 $\mu_B$/$Co^{2+}$ shown in Figure 5 is thus consistent with high-spin $Co^{2+}$ in anatase $TiO_2$ and, importantly, is inconsistent with a low-spin $Co^{2+}$ ground state (S = 1/2, $\mu_{eff}^{s.o.}$ = 1.73 $\mu_B$/$Co^{2+}$).

To demonstrate processability of these colloidal DMS NCs, thin nanocrystalline films were prepared by spin-coating colloidal TOPO-capped $Co^{2+}$:$TiO_2$ NCs onto polycrystalline sapphire substrates. The use of colloidal $Co^{2+}$:$TiO_2$ nanocrystals rather than sol-gel precursors helps to ensure a homogeneous distribution of $Co^{2+}$ ions throughout the film. The resulting films are of high optical quality (Figure 3A, inset) and are largely transparent throughout the visible energy range (Figure 3A). Figure 1(c) shows X-ray diffraction data collected for one such film. The nearly identical X-ray diffraction peak widths for the film and the precursor NCs (NC1, Figure 1(b)) indicate very little NC growth under the relatively mild conditions used to anneal the films (see Experimental section). No evidence of any impurity phase (e.g. cobalt metal) could be detected in either the nanocrystals or the films. As noted previously, however, X-ray diffraction is often too insensitive to detect the impurity phases anticipated in DMSs,[8] and dopant speciation is better evaluated using dopant-specific probes (vide infra). Figure 2B shows an SEM image of a typical film (Film 1). The film appears close to uniform in thickness, but mesoscopic cracking defects are observed that likely arise from calcination of the passivating ligands. Profilometry measurements showed this film to be 500 ± 30 nm thick. All nanocrystalline films were electrically insulating when examined using an in-line four-point probe, presumably due in part to such cracking defects. Very high resistivities have generally been observed in related nanocrystalline films of pure $TiO_2$ nanocrystals.[29]

Magnetic measurements were performed for a series of nanocrystalline and thin-film samples. Figure 6A shows 300 K magnetic hysteresis data measured for two nanocrystal samples: TOPO-capped 3.0 ± 0.1 % $Co^{2+}$:$TiO_2$ NCs (NC1) and 1.3 ± 0.3 % $Co^{2+}$:$TiO_2$ NCs isolated without surface-passivating ligands (NC2). The 300 K magnetization of NC1 follows Curie behavior, consistent with paramagnetic $Co^{2+}$ as concluded from the data in Figure 5. The magnetization data collected for NC2 (Figure 6A) show very weak but distinct evidence of magnetic ordering, with an apparent ferromagnetic saturation moment of $M_s \approx 1.5 \times 10^{-3}$ $\mu_B$/$Co^{2+}$ at 300 K. Although small, this magnetic ordering is of fundamental importance because of the aerobic preparative conditions used, which exclude the possibility that it comes from metallic cobalt. We have previously observed that high-$T_C$ ferromagnetism can be induced in paramagnetic ZnO DMS NCs by room-temperature aggregation in air.[14,30] The change from paramagnetic nanocrystals to ferromagnetic nanocrystalline aggregates was proposed to arise from the increased domain volumes and the introduction of donor-type defects upon aggregation. The data in Figure 6A suggest that a similar phenomenon may also occur for $TiO_2$ DMS NCs, although the magnitude of the effect in this case is too small to be analyzed quantitatively.

Very strong room-temperature ferromagnetism was observed when the paramagnetic TOPO-capped NCs were spin-coated into thin nanocrystalline films. Figure 6B shows 300K magnetization data collected for two films prepared from NC1. Distinct magnetic hystereses are observed in both films, with 300 K ferromagnetic saturation moments of $M_s = 0.3$ $\mu_B/Co^{2+}$ (Film 1) and $M_s = 1.9$ $\mu_B/Co^{2+}$ (Film 2). The 300 K coercivities of both films are $H_c = 55$ Oe (Figure 6B, inset), and the remanences are $M_R = 0.02$ $\mu_B/Co^{2+}$ (7%) and $M_R = 0.09$ $\mu_B/Co^{2+}$ (5%) for Films 1 and 2, respectively. These 300 K saturation moments are approximately two orders of magnitude greater than that of the aggregated NCs (Figure 6A), and are comparable to those obtained for $Co^{2+}:TiO_2$ samples prepared by OPA-MBE ($M_s \approx 1.3$ $\mu_B/Co^{2+}$),[6] PLD ($M_s \approx 1.4$ $\mu_B/Co^{2+}$ and $M_s \approx 0.3$ $\mu_B/Co^{2+}$),[5,31] and RF magnetron co-sputtering ($M_s \approx 0.3$ $\mu_B/Co^{2+}$).[32] In each of these cases, the high-vacuum experimental conditions provided reducing environments that make phase segregation of metallic cobalt possible, and in some cases metallic cobalt nanocrystals were observed,[7,8] leading to the current controversy concerning the origins of ferromagnetism in other $Co^{2+}:TiO_2$ films. In contrast, the magnetic ordering seen in Figure 6B is activated by spin-coat processing of paramagnetic $Co^{2+}:TiO_2$ nanocrystals under aerobic conditions shown to quench the ferromagnetism of cobalt metal nanocrystals,[8,33] and therefore unambiguously arises from a source other than metallic cobalt.

To address the origins of this ferromagnetism, the electronic structures of the cobalt ions in the NCs and the ferromagnetic NC films were further investigated by cobalt K-edge X-ray absorption spectroscopy (XAS). XAS data for freestanding NCs (NC1), spin-coated NC films (Films 1 and 2), and reference samples of metallic cobalt and $CoTiO_3$ are shown in Figure 7A. The 300 K saturation moments of Film 1 and Film 2 (Figure 6B) are 17% and ~110% that of metallic cobalt ($M_s \approx 1.72$ $\mu_B/Co^0$ at 300K[8]). In both films, the presence of such appreciable $Co^0$ would be readily detected in the cobalt K-edge XAS data. The cobalt K-edge spectra of the NCs and NC films are all nearly identical, each showing a sharp 1s → 4p transition edge at 7719.5 eV and a well-resolved 1s → 3d pre-edge feature at 7709.1 eV (Figure 7A inset). These spectra are also nearly identical to the XAS spectrum of a ferromagnetic 5% $Co^{2+}:TiO_2$ thin film grown by OPA-MBE and reported previously.[18] The XAS spectra of the doped TiO$_2$ samples are all very similar to that of CoTiO$_3$, in which cobalt is uniformly in the Co$^{2+}$ oxidation state, confirming a Co$^{2+}$ oxidation state for the TiO$_2$ samples as well. In contrast, the metallic cobalt reference sample shows a broad absorption edge with a prominent pre-edge maximum at 7708.8 eV (Figure 7A, spectrum (e)). From these data, the predominant oxidation state of cobalt in the TiO$_2$ samples is concluded to be Co$^{2+}$, and because of the large saturation moments of these films, these data conclusively rule out cobalt metal impurities as a possible source of the ferromagnetism shown in Figure 6.

Fourier transforms of the extended fine structure (EXAFS) in the XAS data for NC1 and Films 1 and 2 are shown in Figure 6B, in comparison with data collected for a thin film grown by OPA-MBE and for metallic cobalt. Analysis of the EXAFS data yields Co-O bond lengths of 2.07 ± 0.01 Å in NC1 and 2.01 ± 0.01 Å in Films 1 and 2, similar to values obtained for the OPA-MBE thin film (2.01 ± 0.01 Å and 2.04 ± 0.01 Å[18]) and for a powder sample studied previously (2.05 ± 0.01 Å[34]). Notably absent from the EXAFS data is any intensity attributable to the 2.50 Å Co-Co nearest neighbor separation of metallic cobalt,[35] supporting the conclusion drawn above that metallic cobalt is absent in these samples. From the well-resolved pre-edge XAS feature, the XAS edge energy, and the EXAFS data, we conclude that the cobalt in the NCs and the ferromagnetic NC films is best described as Co$^{2+}$ within crystalline anatase TiO$_2$. Taken together with the magnetization data in Figure 6, these results strongly argue for the existence of intrinsic room-temperature ferromagnetism in the diluted magnetic semiconductor Co$^{2+}$:TiO$_2$.

DISCUSSION AND CONCLUSIONS

The principal conclusion of this study is that very strong room-temperature ferromagnetism can be obtained by processing homogeneous colloidal Co$^{2+}$:TiO$_2$ nanocrystals under aerobic conditions that preclude the formation of metallic cobalt. This conclusion strongly supports claims of an intrinsic mechanism for long-range magnetic ordering in this diluted magnetic semiconductor and refutes suggestions that this ferromagnetism arises from metallic cobalt nanoparticles in all instances. As such, this diluted magnetic semiconductor may indeed be suitable for high-temperature semiconductor spintronics applications.

Beyond this central conclusion, the new chemical approach taken here has provided some additional information about magnetic ordering in this material. Although very weak ferromagnetism was observed in $Co^{2+}$:$TiO_2$ nanocrystals aggregated at room temperature, strong ferromagnetism was observed only upon spin-coat processing of the nanocrystals to form the nanocrystalline films. On the one hand, this enhancement of ferromagnetism might be attributed to nanocrystal sintering during the brief anneal process, which would increase the stability of a magnetized domain against magnetization reversal by increasing the magnetocrystalline anisotropy energy.[36] Nanocrystal sintering may also facilitate free-carrier-mediated magnetic ordering by providing enhanced electronic coupling among nanocrystals,[37] if such a mechanism for magnetic ordering were active in this material as has been proposed.[38] The XRD data do not show a significant change in average NC diameter upon annealing, however, suggesting that only limited sintering occurs. We propose that the effect of temperature may instead reflect the presence of an activation barrier to conversion of paramagnetic $Co^{2+}$:$TiO_2$ into a magnetically ordered state. Recent density functional calculations have shown that clustering of substitutional and interstitial cobalt ions within the $TiO_2$ lattice may be able to stabilize ferromagnetism,[39,40] for example, and the brief anneal could activate ferromagnetism by supplying sufficient thermal energy to allow $Co^{2+}$ migration into interstitial sites. Cobalt migration has been observed in $Co^{2+}$:$TiO_2$ thin films heated to 400 °C under low $O_2$ partial pressures ($10^{-6}$ Torr $O_2$),[8] indicating reasonable mobility of this dopant under oxygen-poor conditions. The ferromagnetism observed after vacuum annealing of these films was shown to arise from cobalt metal nanocrystals, however, and furthermore it completely disappeared when the same films were annealed again in air.[8] This observation clearly distinguishes the ferromagnetism of metallic cobalt nanocrystals from that reported here (Figure 6), which increased with annealing in air, and which is attributed to the diluted magnetic semiconductor $Co^{2+}:TiO_2$.

The cobalt K-edge XAS data (Figure 7) show conclusively that annealing does not induce major changes in cobalt oxidation state or cobalt-oxo bond lengths. This observation rules out mechanisms requiring extensive structural reorganization at the $Co^{2+}$ ions themselves. Since the coordination geometries of interstitial dopants in anatase $TiO_2$ are substantially different from those of substitutional dopants, the data in Figure 7B are therefore not consistent with significant migration of cobalt from substitutional to interstitial sites during aerobic anneal. It therefore appears more likely that migration of other defects is responsible for the magnetic phase transition, the most likely candidate being the oxygen vacancies that should accompany $Co^{2+}$ doping to compensate the charge mismatch of the dopant. Activation barriers as low as 20 kcal/mol have been reported for oxygen migration in rutile.[41] We note that the EXAFS data are not consistent with a change in $Co^{2+}$ coordination number with annealing, suggesting that the mobile defect is not in the first coordination sphere of the $Co^{2+}$ ions. Additional experiments to elucidate the microscopic origins of this ferromagnetism are currently in progress. Future experiments will also address the MCD spectroscopy of the ferromagnetic phase in this material.

The conclusion of a high-spin $Co^{2+}$ ground state in the $Co^{2+}:TiO_2$ nanocrystals warrants some discussion. A low-spin ground state has been suggested previously on the basis of the ferromagnetic saturation moments of $Co^{2+}:TiO_2$ thin films. The experimental 300 K saturation moments (e.g. $M_s$ = 1.26 $\mu_B$/Co (ref [6]), 0.32 $\mu_B$/$Co^{2+}$ (ref [5]), 0.94 $\mu_B$/$Co^{2+}$ (ref [42]) were much closer to the spin-only value for low-spin $Co^{2+}$ ($M_s$ = 1 $\mu_B$/$Co^{2+}$) than for high-spin $Co^{2+}$ ($M_s$ = 3 $\mu_B$/$Co^{2+}$). The data in Figure 5 ($\mu_{eff}$(300 K) = 4.2 $\mu_B$/$Co^{2+}$) clearly demonstrate a high-spin $Co^{2+}$ ground state for the paramagnetic phase of $Co^{2+}:TiO_2$, however, a result that is consistent with earlier data for paramagnetic $Co^{2+}$ in bulk $TiO_2$ ($\mu_{eff}$(300 K) = 4.1 $\mu_B$/$Co^{2+}$).[27] One explanation for this apparent discrepancy may be that the paramagnetic-to-ferromagnetic phase transition of $Co^{2+}:TiO_2$ is coupled to a $Co^{2+}$ ground-state spin crossover. The data in Figure 7 show a contraction of $Co^{2+}$-oxo bonds by ca. 0.06Å upon forming the films, and such a contraction would be consistent with a high- to low-spin conversion. This possibility seems unlikely in view of the high saturation moments of the ferromagnetic films measured here ($M_s$ = 1.9 $\mu_B$/$Co^{2+}$ at 300 K, Figure 6), however. Related to this is the possibility that different $Co^{2+}$ dopant sites may have different spin states, as suggested by *ab initio* calculations,[38,39] and that variations in processing procedures may yield different ratios of site occupancies. This, too, seems rather unlikely to be able to directly account for such a large range of saturation moments. In our ferromagnetic nanocrystalline films, the range of saturation moments is at least in part due to the incomplete conversion of paramagnetic $Co^{2+}$:$TiO_2$ into the ferromagnetic phase. This results in the presence of substantial residual paramagnetic $Co^{2+}$, and potentially some $Co^{2+}$ rendered magnetically silent through antiferromagnetic superexchange. Because these forms of $Co^{2+}$ do not saturate at the low fields used to measure the ferromagnetic saturation moments, the resulting $M_s$ values provide only lower limits for the fully ferromagnetically ordered materials. If this were also true for the materials from other laboratories, it would explain the apparent discrepancy in $Co^{2+}$ spin states.

In summary, we have presented the synthesis, absorption spectra, 300K magnetic susceptibility, and cobalt K-edge X-ray absorption and EXAFS data of $Co^{2+}$:$TiO_2$ NCs and nanocrystalline films. Absorption, MCD, and magnetic susceptibility data are consistent with $Co^{2+}$ ions occupying the $D_{2d}$ anatase cation substitution sites and having a $^4E(^4T_{2g})$ low-symmetry ground state. Strongly ferromagnetic thin films of $Co^{2+}$:$TiO_2$ ($M_s$ = 1.9 $\mu_B$/$Co^{2+}$ at 300 K) were prepared by spin-coat processing of these nanocrystals under aerobic conditions that preclude the formation of metallic cobalt. Cobalt K-edge X-ray absorption spectra show that the majority of the cobalt is in the $Co^{2+}$ oxidation state and coordinated by lattice oxos in both the NCs and the ferromagnetic nanocrystalline films. These data thus rule out the presence of sufficient cobalt metal to explain the strong ferromagnetism, and instead provide strong experimental support for the existence of intrinsic room-temperature ferromagnetism in cobalt-doped $TiO_2$. The preparation of this DMS in the form of colloidal nanocrystals and its compatibility with oxidative processing conditions raise the possibility of using such DMS colloids as building blocks for the assembly of new high-$T_C$ ferromagnetic semiconductor nanostructures by bottom-up techniques such as soft lithography or self assembly.

**Acknowledgment.** Financial support from the NSF (DMR-0239325 and ECS-0224138), Research Corporation (RI0832), ACS-PRF (37502-G), and the Semiconductor Research Corporation (2002-RJ-1051G) is gratefully acknowledged. J.D.B. thanks the UW/PNNL Joint Institute for Nanoscience for a post-doctoral fellowship. TEM data were collected at the EMSL (PNNL), a national scientific user facility sponsored by the DOE. Dr. Chongmin Wang (PNNL) and Nick Norberg (UW) are thanked for valuable TEM assistance. We thank Prof. Jesper Bendix (U. Copenhagen) for generously providing the LIGFIELD software used to calculate the energies of Figure 4. D.R.G. is a Cottrell Scholar of the Research Corporation.

**Supporting Information Available:** Overview low-resolution TEM image of TOPO-capped 3% $Co^{2+}$:$TiO_2$ NCs (NC1). This material is available free of charge via the Internet at http://pubs.acs.org.

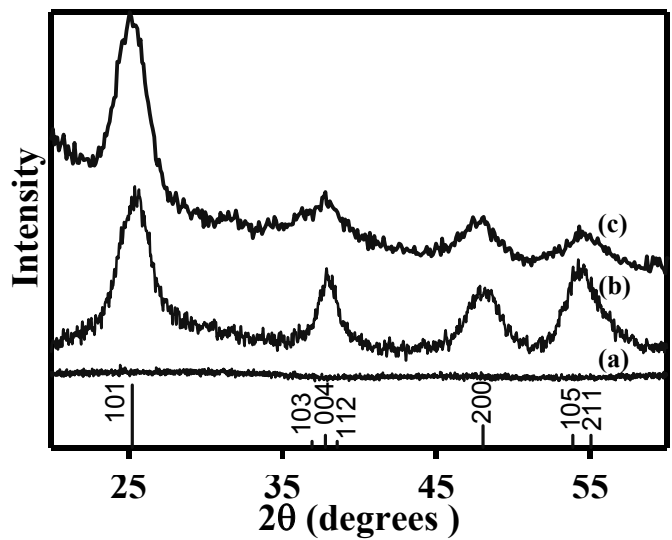

**Figure 1.** X-ray powder diffraction: (a) amorphous precursor, (b) TOPO-capped 3% $Co^{2+}$:$TiO_2$ NCs (NC1), (c) spin-coated 3% $Co^{2+}$:$TiO_2$ NC film. Bragg peak positions, Miller indices, and powder intensities for anatase $TiO_2$ are included for reference.

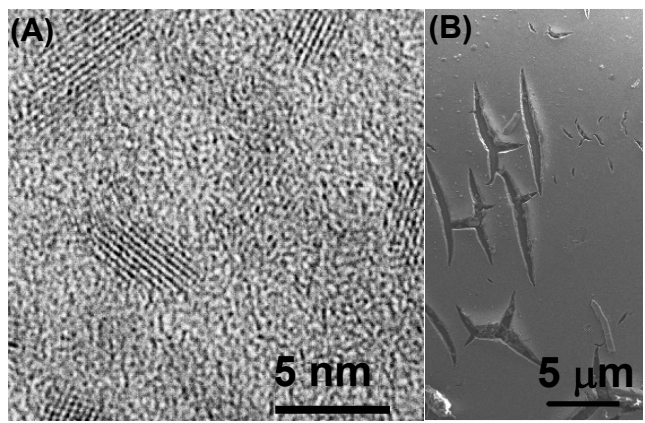

**Figure 2. (A)** High-resolution TEM image of TOPO-capped 3% $Co^{2+}$:$TiO_2$ NCs (NC1) **(B)** SEM image of a spin-coated nanocrystalline film (Film 1).

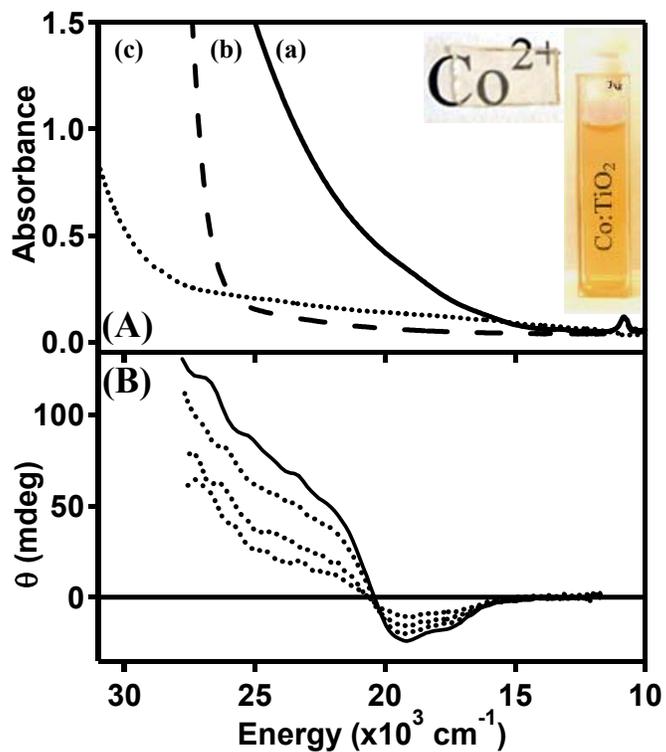

**Figure 3.** **(A)** 300 K absorption spectra: (a) TOPO-capped 3% $Co^{2+}$:$TiO_2$ NCs (NC1) in toluene, (b) undoped $TiO_2$ NCs in toluene, (c) spin-coated film of 3% $Co^{2+}$:$TiO_2$ NCs (Film 1). Inset: Photos of NC1 suspension and Film 1, with lettering behind each to illustrate transparency. **(B)** 5.5 T variable-temperature (5 (solid), 10, 20, 40 K) MCD spectra of 1.3% $Co^{2+}$:$TiO_2$ NCs.

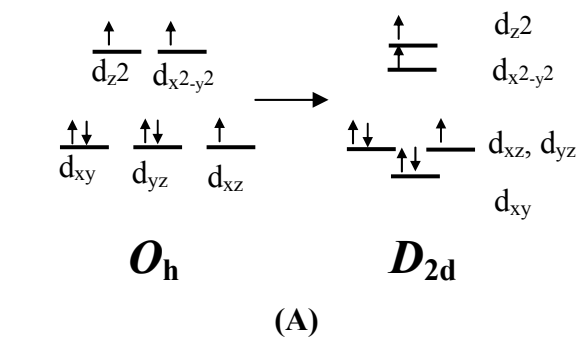

(A)

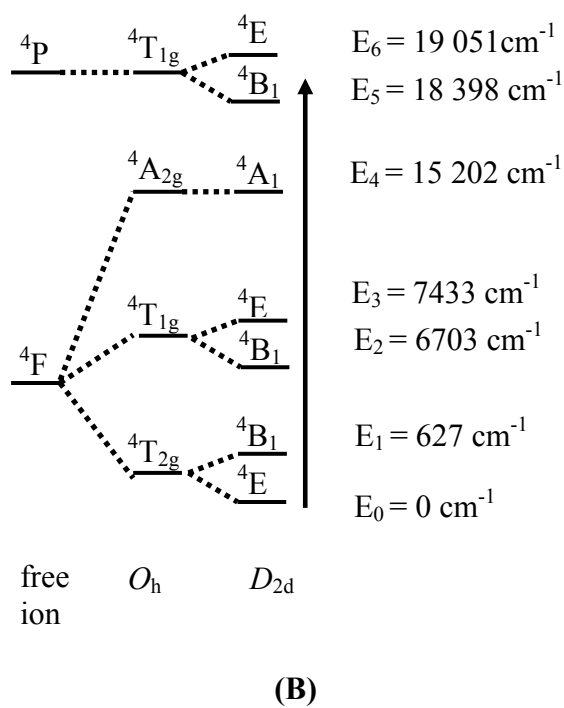

(B)

**Figure 4.** **(A)** Schematic d-orbital splitting diagram for $Co^{2+}$ in $O_h$ and $D_{2d}$ (anatase) geometries. **(B)** Term diagram showing splitting of the $Co^{2+}$ free ion terms in $O_h$ and $D_{2d}$ (anatase) ligand fields. The estimated energies of the terms for $Co^{2+}$ in anatase $TiO_2$ are shown on the right (see text). The vertical arrow indicates the excitations observed in absorption and MCD spectra (Figure 3).

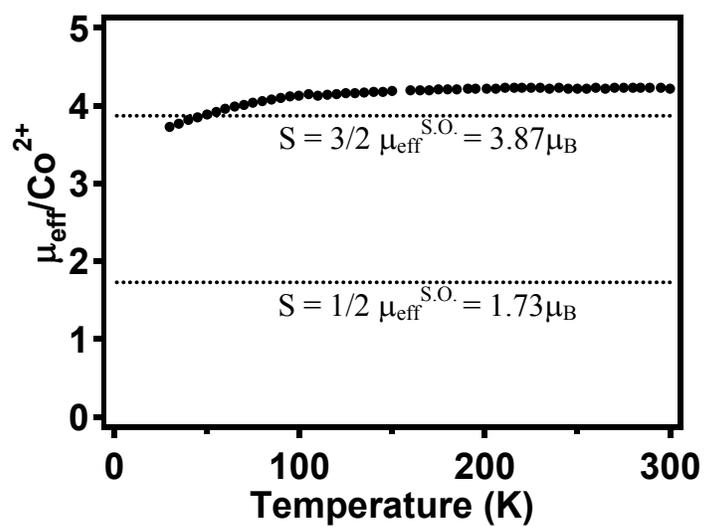

**Figure 5.** Temperature dependence of the effective magnetic moment ($\mu_{eff}$) measured for $3.0 \pm 0.1\ \%$ $Co^{2+}$:$TiO_2$ NCs (NC1). H = 1000 Oe. Spin-only $\mu_{eff}$ values for high- and low-spin $Co^{2+}$ are shown as dashed lines.

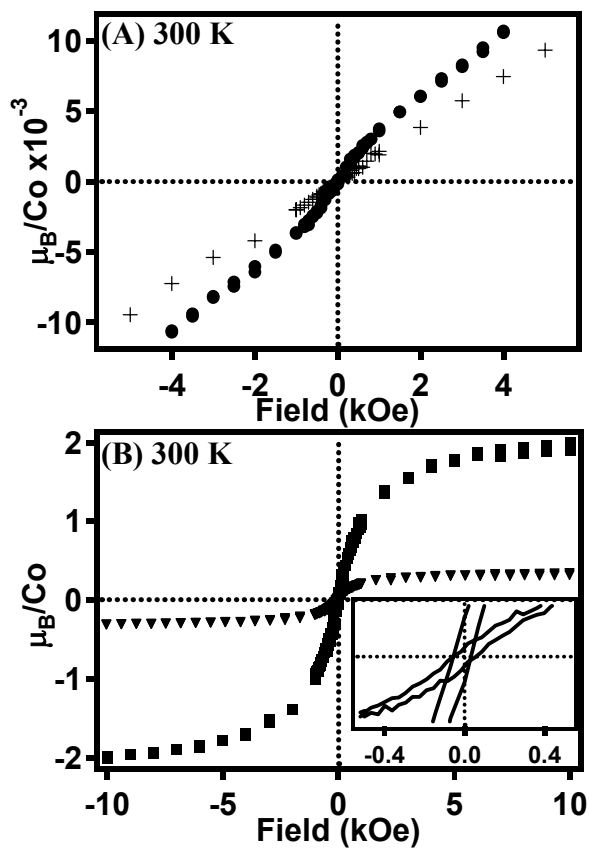

**Figure 6.** 300 K magnetization data: **(A)** (+, NC1) TOPO-capped 3.0 ± 0.1 % $Co^{2+}$:$TiO_2$ and (•, NC2) 1.3 ± 0.3 % $Co^{2+}$:$TiO_2$ nanocrystals. **(B)** (▼) Film 1 and (■) Film 2 (spin-coated films of NC1). The inset shows the coercivities of the films. Note the different axis scales. All data have been corrected for diamagnetism.

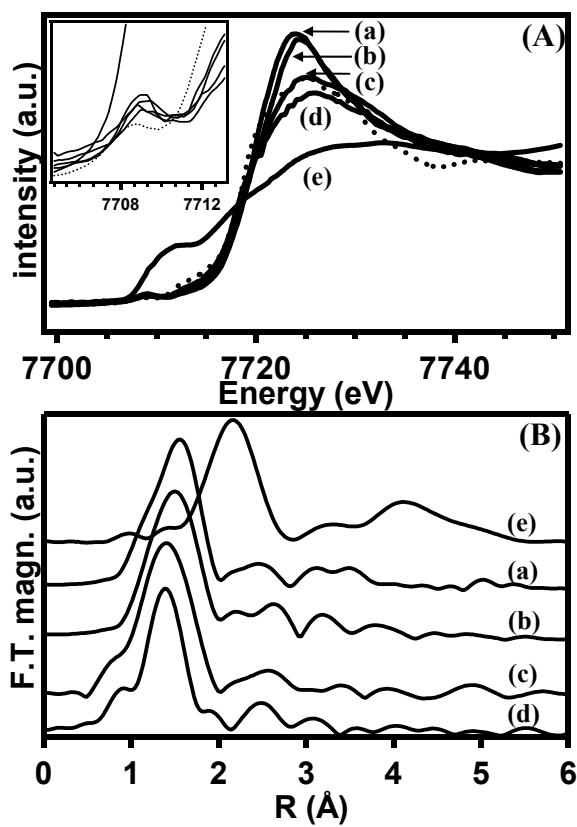

**Figure 7. (A)** Cobalt K-shell X-ray absorption spectra for (a) NC1 (freestanding 3% $Co^{2+}$:$TiO_2$ nanocrystals), (b) Film 2, (c) Film 1, (d) 5% $Co^{2+}$:$TiO_2$ film grown by OPA-MBE (ref 18), (e) Co metal foil. The dashed line is the $Co^{2+}$ reference compound, $CoTiO_3$. **(B)** Fourier transform magnitude of the Co K-edge EXAFS data. The lettering scheme is the same as in (A).

**Table of contents graphic:**

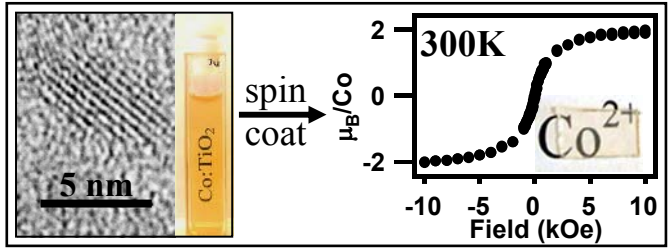